\documentclass[conference]{IEEEtran}
\usepackage{wrapfig}
\usepackage{multicol}
\usepackage{lipsum}
\usepackage{amsmath}
\usepackage{mathtools}
\usepackage{caption}
\usepackage{subcaption}
\usepackage{xcolor}
\usepackage{url}

\usepackage[ruled,vlined]{algorithm2e}

\usepackage{flushend}
\usepackage{balance}

\newcommand{\ie}{\emph{i.e.}\xspace}

\newcommand{\Comment}[1]{}

\newcommand{\approachName}{ASC$_{\text{RL}}$}
\newcommand{\iap}{\emph{I}-App}
\newcommand{\iaps}{\emph{I}-Apps}

\author{\IEEEauthorblockN{Jinming Xing, Muhammad Shahzad}
\IEEEauthorblockA{
\textit{North Carolina State University, USA}}
}

\begin{document}
\title{A Reinforcement Learning Framework for\\Application-Specific TCP Congestion-Control}
\maketitle
{
\sloppy
\begin{abstract}
The Congestion Control (CC) module plays a critical role in the Transmission Control Protocol (TCP), ensuring the stability and efficiency of network data transmission.
%
The CC approaches that are commonly used these days employ heuristics-based rules to adjust the sending rate.
It is well known that due to their heuristics-based nature, these approaches are not only unable to adapt to changing network conditions but are also agnostic to the diverse requirements that different applications often have.
Lately, we have seen several learning-based CC approaches that do not rely on fixed heuristics-based rules and thus can adapt to changing network conditions.
Unfortunately, existing learning-based CC approaches have also not been designed to take application requirements into account.
Prior heuristics-based as well as learning-based CC approaches focus on achieving a singular objective, which is often to maximize throughput, even though a lot of applications care more about latency, packet losses, jitter, and/or different combinations of various network metrics.
Motivated by this, we propose a Deep Reinforcement Learning (DRL) based CC framework, namely \approachName, which allows any application to specify any arbitrary objectives that the network traffic of that application should achieve and is able to swiftly adapt to the changes in the objectives of the applications as well as to the changes in the network conditions.
Our \approachName~framework further employs a client-server architecture that serves two purposes: 1) it makes \approachName~highly scalable in terms of the arrival and departure of TCP connections, and 2) it makes \approachName~very lightweight for the nodes maintaining the TCP connections.
We implemented and extensively evaluated \approachName~in a variety of settings.
Our results show that it can not only achieve various objectives but also outperforms prior approaches even in the specific objectives that those approaches were designed to achieve.
We will open-source the code and data used in the evaluation of \approachName~on the acceptance of this paper.
%
\end{abstract}

\section{Introduction}
%
Congestion control algorithm's (which are mostly used by TCP but lately also by other transport and application protocols, such as QUIC and webRTC) are the foundational protocols that enable the Internet to successfully transport the data of billions of devices without getting choked.
Unfortunately, the existing approaches to congestion control are not necessarily the most optimal for the networked ecosystems of today, which consist of numerous types of devices and myriads of diverse Internet applications (subsequently referred to as \iaps) that communicate over the Internet.
In particular, there are two fundamental limitations that existing approaches have.
But before we describe them, let us first quickly overview the current landscape of congestion control methodologies.

Existing congestion control methodologies can be broadly categorized as \emph{heuristics-based} and \emph{learning-based}.
The heuristics-based approaches use fixed-rules that typically rely on round trip delay and/or packet loss information.
The well-known examples of such approaches include New Reno \cite{grieco2004performance}, Cubic \cite{Ha2008}, DCTCP \cite{Alizadeh2010}, and BBR \cite{scholz2018towards}.
The learning-based approaches, as the name suggests, learn over time the best way to respond to current network conditions to achieve an \emph{objective}.
For all existing learning-based approaches, the \emph{objective} typically has been to achieve the highest throughput.
Examples of existing such approaches include QTCP \cite{li2018qtcp}, DRL-CC \cite{xu2019experience}, TCP-Drinc \cite{xiao2019tcp}, and SmartCC \cite{li2019smartcc}.
Learning-based approaches are newer compared to the heuristics-based approaches and hold the promise to provide tailored services to a diverse set of \iaps~with a diverse set of requirements running on a diverse set of networked devices.

\subsection{Limitations of Prior Art}
\label{subsec:LimitationsofPriorArt}
While the work on congestion control has been plenty, prior approaches suffer from either or both of the following two fundamental limitations.

\vspace{0.03in}
\noindent\textbf{Application-agnostic:} 
Different \iaps~have different needs.
For example, file transfer needs high throughput \cite{yildirim2015application}, audio calls need low latency \cite{shaw2016survey}, cloud streamed games require low latency as well as small jitter \cite{shahzad2014noise}, and so on.
Unfortunately, as heuristics-based approaches rely on fixed rules, they cannot take into account such requirements of a diverse set of \iaps~simultaneously.
Similarly, all existing learning based approaches fall short as they focus on a singular objective, such as maximizing throughput, and assume that all \iaps~have that same objective.

The objectives of many \iaps~of today change during different phases of their operation.
For example, during the initial buffering phase, video streaming applications require high throughput to quickly fill the buffer to a sufficient level so that video playback starts right away.
During the next playback phase, video streaming requires minimal jitter and packet loss to maintain smooth playback of the stream.
During any re-buffering phase, video streaming again requires high throughput.
Similarly, during web browsing, whenever a user clicks a link, low latency and high throughput are needed to minimize page load time.
Afterwards, as the user interacts with the loaded page, throughput becomes less important and only low latency is needed to facilitate smooth interactions.
Yet another example is video conferencing, where signaling/connection setup phase is sensitive to latency and jitter, video/audio streaming phase is additionally sensitive to packet loss, screen/content sharing phase is sensitive to throughput and latency, and remote screen control phase is sensitive to latency.
A newer \iap, Cloud-streamed gaming, requires either low latency (to display immediate response when user provides inputs) or high throughput (to transfer video frames from cloud to the user's device) and frequently switches between these two requirements as the user plays the game.
There are numerous other such examples of \iaps~whose objectives change during different phases of their operations.
Unfortunately, neither the heuristics-based nor the existing learning-based approaches can handle such changes in objectives.

\vspace{0.03in}
\noindent\textbf{Computational demand:} 
While heuristics based approaches are computationally inexpensive, the learning based approaches are not because they need to be trained and retrained to adapt to the latest network conditions.
%
%
Existing learning-based approaches are computationally demanding not only for the low-powered devices but also for typical computers/laptops.
This is because typical computers have multiple ongoing connections, and this training and retraining often needs to be done separately for each TCP connection.
%

\vspace{-0.05in}
\subsection{Problem Statement}
\label{subsec:ProblemStatement}
Our goal is to develop a congestion control framework that allows any \iap~to specify any arbitrary objectives that the network traffic of that \iap~should achieve, such as maximize throughput, minimize latency/jitter, optimize any other network metric, or any combination of these objectives.
The framework should further meet the following three requirements:
(1) adapt to changes in the objectives of the \iap,
(2) require minimal computational capability from the node running the \iap, 
and
(3) be scalable in terms of swiftly handling new and terminating TCP connections.


\vspace{-0.05in}
\subsection{Proposed Approach}
\label{subsec:ProposedApproach}
We present \approachName, a \underline{r}einforcement \underline{l}earning \underline{c}ongestion control framework that achieves any \underline{a}pplication-\underline{s}pecified objectives while meeting the three requirements listed above and addressing the two limitations of prior art.
Logically, \approachName~performs two functions: learning and inference.
Inference is computationally cheap and is performed on the node running the \iap(s).
Learning, however, can be computationally heavy and, thus, \approachName~performs it on a dedicated server.
Onwards, we will refer to the nodes performing the inference as CC-clients (\ie, congestion control clients) and the server performing the learning as CC-server.
%
%


Fig. \ref{fig:framework} provides a high-level overview of \approachName.
For any given set of CC-clients on a non-wide area network (such as all the devices in a house, in a subnet of an enterprise network, or in a data center rack/room), the CC-server maintains a \emph{central-model} that continuously keeps learning the characteristics of the traffic
of the CC-clients located within this non-wide area network.
%
%
For each \iap~on each CC-client, the CC-server further maintains an individualized \emph{sub-model}.
%
Each sub-model is derived from the central-model and is tailored to achieve the objectives that the \iap~corresponding to that sub-model has specified.
%
\begin{figure*}[htbp]
\centering
    \includegraphics[width=0.8\linewidth]{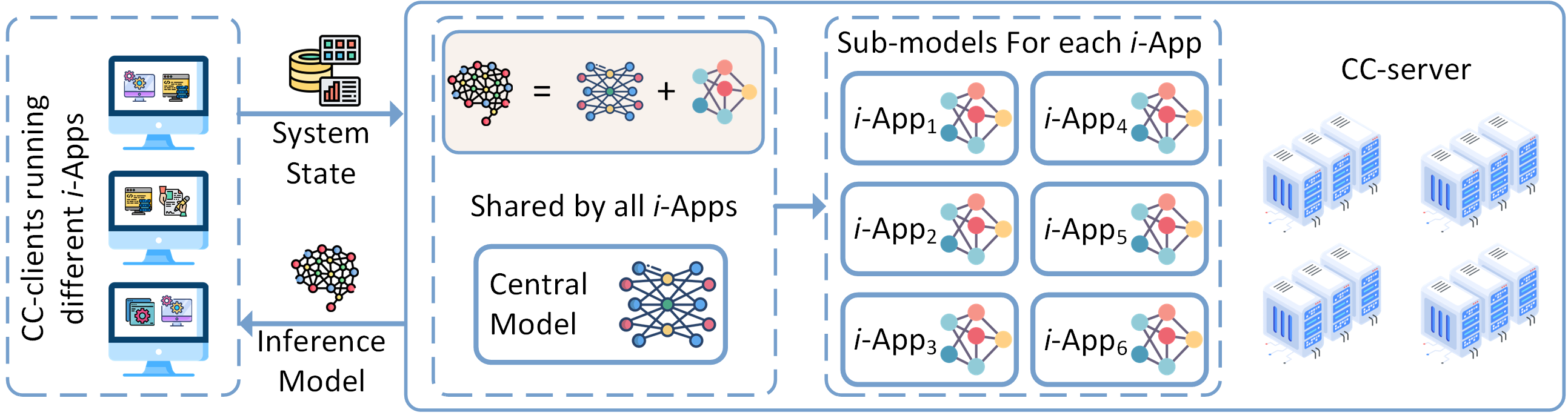}
\vspace{-0.05in}
    \caption{A high-level overview of \approachName}
\vspace{-0.23in}
    \label{fig:framework}
\end{figure*}

Each CC-client periodically transmits some network statistics, called system state (details on the specifics of system state later), of each \iap~running on it to the CC-server.
Every time the CC-server receives the new-system state of any \iap~from a CC-client, it updates the central-model as well as the sub-model of that \iap~using reinforcement learning, and sends the updated inference-model back to the CC-client to be used for that \iap.
Any CC-client receives inference-models only for the \iaps~executing on it.
%
The CC-client then uses these inference-models to set the congestion window (cwnd)
sizes for the TCP connections of the corresponding \iaps.
The use of these individualized inference-models ensures that the \iaps~achieve their specific objectives.
The duration that the CC-client waits for while collecting system state before sending it to the CC-server and receiving the updated model from the CC-server warrants attention.
A shorter updating interval results in increased resource consumption due to frequent communication with the server, while a longer interval hinders model's performance by lacking latest information.
In the evaluation section, we will thoroughly analyze the impact of updating interval on the network overhead and on \approachName's ability to achieve the specific objectives of the \iaps.
%
%


\vspace{0.03in}
\noindent\textbf{Adaptability:} 
Our \approachName~framework can handle the scenarios where any given \iap~updates/changes its objectives.
When this happens, CC-server retrains the sub-model corresponding to that \iap, but not from scratch.
\approachName~introduces a fast retraining mechanism (FRM), where instead of retraining from scratch, CC-server first determines whether any other \iap~communicating through the same bottleneck link has a similar objective.
If it identifies such an \iap, it uses the sub-model of that \iap~as the starting model and adapts it for the \iap~that just changed its objectives.
This approach minimizes the need for extensive fine-tuning, thereby achieving significantly faster retraining.
If the CC-server does not find another \iap~with similar objective, it organically fine-tunes the sub-model of the \iap~that just changed its objective.
Note that the fine-tuning is still a quick process because the sub-model is derived from a well-trained central-model.
%

\vspace{0.03in}
\noindent\textbf{Minimal Computational Requirements:} 
In \approachName, as inference is computationally cheap, it is performed on the CC-client running the \iap(s).
Learning is computationally complex, and thus performed on a dedicated CC-server.
%
This deployment model makes \approachName~amenable even for low-powered devices.
In household settings, the CC-server could be a typical computer in the house, or a modern modem/router installed by the ISP in the house.
The CC-server could even be provided as a cloud based service (we have not evaluated \approachName~with such a deployment of CC-server though).
%
In enterprise settings, the CC-server could be deployed along with a DHCP server, a web server, an authoritative DNS server, an internet proxy, or any other such server.
%
%
Similarly, in data centers, the CC-server could be deployed on one or more dedicated servers situated within that data center.

\vspace{0.03in}
\noindent\textbf{Scalability:} 
The \approachName~framework is also highly scalable.
To handle a new \iap, the CC-server simply starts a new sub-model for that \iap, uses the FRM approach, if possible, to quickly tune it.
It then sends the trained inference-model to the CC-client running the new \iap.
If an \iap~terminates its TCP connection(s), the CC-server simply terminates the corresponding sub-model.
%

\subsection{Key Contributions}
In this paper, we make the following five key contributions:
%
\begin{enumerate}
\item We propose the \approachName~framework, a learning-based congestion control approach that decouples learning from inference, making it lightweight and thus usable on commodity devices.
%
\item  We have designed \approachName~such that, for the first time, different \iaps~can achieve different objectives on the same network.
%
\item We introduce FRM, which makes it practical for \iaps~to change and quickly achieve their objectives.
%
\item \approachName~is highly scalable in terms of swiftly adding new \iaps~and removing those that no longer send data.
%
\item We extensively evaluated \approachName~on multiple topologies and showed that it not only achieves diverse objectives but also outperforms prior approaches, even in the specific objectives those approaches were designed for.
\end{enumerate}
We will release the source code and data used in the evaluation and comparison of \approachName~on the acceptance of this paper.

\section{Related Work}
\textbf{Heuristics-Based Congestion Control}.
Heuristic-based congestion control methods adjust the congestion window by utilizing predefined condition-action tables.
In these methods, a TCP sender typically uses packet losses and round trip time (RTT) as indicators of the network conditions.
TCP CUBIC \cite{Ha2008} employs a cubic function of time since the most recent packet loss to adjust its \texttt{cwnd}.
TCP-Illinois \cite{liu2006} aims to serve the needs of long-distance networks and relies on both packet losses and RTT to learn network conditions.
FAST-TCP \cite{jin2004fast} aims to maintain a constant occupancy of the queue based on the RTT difference.
%
%
%
DCTCP \cite{Alizadeh2010} focuses on data center networks and employs explicit congestion notification to mitigate congestion.
%

\noindent\textbf{Learning-Based Congestion Control}. 
Unlike heuristics-based approaches, learning-based methods have the ability to learn policies from the environment rather than relying on predefined rules \cite{Sivaraman2015,Winstein2013}.
Several works belonging to this category treat congestion control as an optimization problem and define utility functions \cite{Sivaraman2015,Winstein2013,Dong2015}.
%
%
Lately, the potential of reinforcement learning in congestion control has also been explored \cite{Jiang2021,li2018qtcp,Jay2019}.
Li et. al. explored Q-Learning \cite{li2018qtcp}, while Tessler et al. investigated the policy gradient method \cite{Tessler2022} in data center networks.
%
He et al. \cite{He2021b} introduced RMTC, capable of achieving real-time congestion control and multipath scheduling.
%
In \cite{He2021b}, multi-agent deep reinforcement learning and self-attention mechanisms were adopted for multipath congestion control.
There are several other approaches that use similar learning-based techniques to address congestion, such as \cite{zhang2020machine, Jay2019, luo2025rank}.
Interested readers can find comparisons and analyses of the prominent learning-based congestion control methods in \cite{Jiang2021,Arianpoo2016,Wang2018}.
Unfortunately, all prior learning-based approaches suffer from the key limitations mentioned earlier in Sec. \ref{subsec:LimitationsofPriorArt}.
These include their inability to handle diverse optimization objectives from diverse \iaps, their inability to handle changes in the optimization objectives of the \iaps, and their high computational demand from clients executing these \iaps.
%

\vspace{0.05in}
\noindent\textbf{Learning-Based Network Optimization}.
While not directly related, for the sake of completeness, we also mention some approaches that have used learning-based techniques to solve various network optimization problems.
These problems include network data slicing \cite{Wang2019b,Liu2021}, load balancing \cite{Zheng2023}, and resource management \cite{xing2025netsight}.
Luo et al. explored multipath data scheduling in \cite{Luo2019}, while Chinchali et al. \cite{Chinchali2018} proposed a self-adaptive RL model for scheduling IoT traffic.
Xie et al. \cite{Xie2020} introduced a DRL-based method for adjusting the initial congestion window in 5G mobile networks.
In \cite{nie2019}, 
Nie et al. proposed TCP-RL, a DRL model that co-optimizes the initial window and congestion window.
%
%
Li et al. \cite{Li2020} utilized GNNs to model network traffic, optimize flow routing, and perform topology management, resulting in reduced flow completion time.
In \cite{Almasan2022b}, Almasan et al. investigated the combination of DRL with GNNs and demonstrated its potential through a routing optimization problem. For further details on GNNs in computer networks, refer to \cite{Jiang2022,xing2024network}.

\section{\approachName~-- Technical Details}
%
We described the overall architecture of \approachName~in Sec. \ref{subsec:ProposedApproach}.
As the learning approach that \approachName~uses is the deep reinforcement learning (DRL), we first provide a quick primer of DRL.
As we will see during the primer, the four components that we need to carefully design when using DRL include the System State, the Reward Function, the Action Space, and the Neural-Network Architecture.
Therefore, after providing the primer, we will discuss how we design these four components of the DRL used in \approachName.

\vspace{-0.03in}
\subsection{DRL Primer}
\label{subsec:DRLPrimer}
Conceptually, a DRL model aligns with a Markov Decision Process (MDP), comprising the following five components:
%

\begin{itemize}
    \item\textbf{Set of States, $S$}, also referred to as System State, which quantifies the current status of the system under consideration and serves as the input to the DRL model.
    \item \textbf{Set of Actions, $A$}, also referred to as Action Space, which is comprised of all the actions that the DRL model can undertake for the given state.
\vspace{0.03in}
    \item \textbf{Reward Function, $R(s,a)$}, which provides the theoretical formulation of quantifying the reward for executing action $a\in A$ when in state $s\in S$.
\vspace{0.03in}
    \item \textbf{Transition Function, $T(s'|s,a)$}, which quantifies the probability of the system transitioning into state $s'\in S$ when action $a\in A$ is taken while the system was in state $s\in S$.
\vspace{0.03in}
    \item \textbf{Discount Factor, $\gamma\in[0, 1]$}, which determines the present value of future rewards. 
\end{itemize}
The primary objective of DRL is to maximize the cumulative reward.
Let $T$ represent the total time steps and $r_{t+1}$ denote the current reward at time $t+1$.
Then the cumulative reward $R$ is calculated as:
$
    R=\sum_{t=0}^{T-1}\gamma^tr_{t+1}
$.
Note that the size of the time-step is a tunable parameter and can range from a few milliseconds to several seconds, depending on the computational resources available at the CC-server.
The smaller the time step, the faster the new reward calculation and thus faster potential convergence of the model at the expense of higher computational requirements at the CC-server.
To achieve maximal cumulative reward, the DRL model necessitates a stochastic policy function $\pi$.
This function determines the action that the model would take given the state of the system.
Formally, $\pi$ is a mapping from states to a probability distribution over actions, denoted as $\pi\rightarrow p(A|S)$.
The DRL model aims to discover the optimal value of $\pi$, denoted with $\pi^*$, to maximize the cumulative reward.
This is often expressed as:
\begin{equation}
    \pi^*=\arg\max_\pi E[R|\pi]
    \label{policy pi}
\end{equation}

The most recent DRL approaches can be classified into two high-level categories: on-policy and off-policy.
In this paper, we concentrate on off-policy methods.
Among these methods, the Soft Actor-Critic (SAC) \cite{Haarnoja2018}, a fundamental model in DRL, serves as our base model.
SAC belongs to the family of actor-critic methods, where the actor neural-network learns the stochastic policy $\pi$
while the critic neural-network estimates Q-function, which quantifies the expected cumulative reward that can be obtained by taking action $a\in A$ while in state $s\in S$.
SAC is advantageous in that it introduced the entropy term to the objective function, enhancing exploration and preventing premature convergence.

From the discussion above, it is evident that the information encapsulated in the system state, the reward, the action space, and the neural network, all play pivotal roles within the \approachName~framework.
Next, we first describe these four components in detail.
After that, we discuss a fast retraining mechanism that \approachName~uses and then conclude the section with a discussion on \approachName's scalability.

\subsection{System-State}
\label{subsec:SystemState}
After extensive investigation, we identified 12 attributes, collectively referred to as system-state, which comprehensively characterize the traffic behavior in a networked-system in the context of TCP congestion control.
These attributes are what any given CC-client collects for each \iap~and after each time step, uses them as an input to the model it received from the CC-server for the given \iap~to obtain the next value of congestion window, \texttt{cwnd}, for the TCP connection of that \iap.
Each CC-client sends the system-state, \ie, the values of these 12 attributes, periodically to the CC-server for reinforcement learning and updating the models to address the latest conditions.
%

\vspace{0.04in}
\noindent\textbf{Timestamp (1 value)}: Timestamps allow the model to learn any temporal patterns that appear repeatedly in network traffic.

\vspace{0.06in}
\noindent\textbf{Congestion Window Size (2 values)}: \texttt{cwnd} provides information about how much data the \iap~tried to transmit.
As \texttt{cwnd} can change over time, relying on the latest \texttt{cwnd} value is insufficient.
In \approachName, for any given \iap, in addition to the current \texttt{cwnd}, the CC-client maintains an exponentially weighted moving average of \texttt{cwnd} as well for that \iap~as part of that \iap's system state.
Formally, at time $t$, this average, $\overline{C}_{t}$, is computed as:
    \begin{equation}
      \overline{C}_{t}={\frac{\sum_{i=0}^K\zeta^i \texttt{cwnd}_{t-i}}{\sum_{i=0}^K\zeta^i}}
      \label{mCwnd}
    \end{equation}
where $\zeta\in(0,1)$ denotes the discount factor, and $K$ specifies how far in the past do we take the \texttt{cwnd} values from.
For clarification, the term $t-i$ in the expression above refers to the value of \texttt{cwnd} $i$ time-steps prior to the current time-step.

\vspace{0.04in}
\noindent\textbf{Acknowledged Segments (2 values)}:
Whenever the receiver receives a segment from the sender, it sends the acknowledgment (\texttt{ACK}) of that segment back to the sender.
The number of segments \texttt{ACK}'d during any time-step directly correlates with the quality of the path between the sender and receiver, and is thus very useful for the DRL model.
Similar to \texttt{cwnd}, for any given \iap, in addition to the number of \texttt{ACK}s received in the current time step, the CC-client maintains an exponentially weighted moving average of the number of segments \texttt{ACK}'d in the past $K$ time steps as part of the system state.

\vspace{0.04in}
\noindent\textbf{Bytes In-flight (2 values)}:
Bytes in-flight is a measure of the amount of data that is currently in transit, \ie, the data that has been sent but not yet \texttt{ACK}'d.
This metric offers the DRL model insights into the volume of data sent and the current congestion level of the network from a broader perspective.
In addition to the current bytes in-flight, the CC-client maintain their exponentially weighted moving average as well in the past $K$ time steps.

\vspace{0.04in}
\noindent\textbf{RTT (2 values)}:
Round-trip time (RTT) is the time it takes for a segment to go from the sender to the receiver node and its \texttt{ACK} to go from the receiver back to the sender.
RTT is a crucial metric that quantifies the congestion on the path from the sender to the receiver, and is thus imperative for the DRL model to have.
Similar to other metrics, a CC-client maintains the exponentially weighted value of RTT for any given \iap.

\vspace{0.04in}
\noindent\textbf{Throughput (2 values)}:
Throughput and its exponentially weighted version represent the rate at which data is successfully delivered by the network from the sender to the receiver.
They encompass information regarding both congestion and packet error rates and thus provide the DRL model with a comprehensive overview of the path from sender to receiver.
%



\vspace{0.04in}
\noindent\textbf{Optimization Objective (1 value)}: 
This value identifies exactly which optimization objective any given \iap~wants to achieve going forward.
As soon as an \iap~with existing TCP connection changes its optimization objective, or as soon as an \iap~establishes a new TCP connection with a certain optimization objective, it informs the CC-server of that objective.
%

%

Note that the goal of any optimization objective is always to either maximize (such as maximize throughput) or to minimize (such as minimize latency, loss, and/or jitter).
In \approachName, we require that an \iap~always specify its objective as a maximization objective.
Thus, if an \iap~needs to actually minimize some metric (such as latency) or a combination of metrics, \approachName~requires it to simply put a negative sign in front of the objective to make it a maximization objective.

\vspace{-0.03in}
\subsection{Reward Function}
\vspace{-0.03in}
\label{subsec:RewardFunction}
%
%
A carefully crafted definition of the reward function is crucial to ensure that any given \iap~achieves its desired objective while being fair with other \iaps~running on the same or even different CC-clients in terms of using the bandwidth of the bottleneck link.
The reward function should further be oblivious to the actual expressions of the objectives of various \iaps~to ensure that it can handle diverse objectives across different \iaps.

Given these requirements, let $o^i$ and $\overline{o^i}$ represent the most recent and weighted objective values received by the CC-server for \iap~$i$.
Similarly, let $\overline{\tau^i}$ represent the exponentially weighted moving average of the throughput achieved by \iap~$i$.
The CC-server receives $\overline{\tau^i}$ as part of the system state for all \iaps~periodically from the CC-clients.
Let $R_i$ represent the reward value for \iap~$i$ at that time step.
Our reward function, which satisfies the requirements listed in the last paragraph, is given by the following expression.
%
\begin{equation}
    R^i =\begin{cases} +1 & \text{if}\ S^i>0 \\
              -1 & \text{if}\ S^i<0 \\
              0  & \text{otherwise}
    \end{cases}
\end{equation}
where
\small
\begin{equation*}
        S^i = \alpha\times \underline{\min\big(C_\tau-\max_j(|\overline{\tau^i}-\overline{\tau^j}|),0\big)}+
        \left(1-\alpha \right)\times\underline{\underline{\lambda\big\{o^i-\overline{o^i}\big\}}}
\end{equation*}
\normalsize
where $\alpha\in[0,1]$ is a constant that we have introduced to offer the flexibility to quantify the desired trade-off between fairness and achieving the objective.
The term underlined with a single line enables the DRL model to ensure that \iap~$i$ shares the bandwidth of any bottleneck link in the network fairly with other \iaps.
The constant $C_\tau$ enables a network manager to quantify how much difference in the bandwidth captured by different \iaps~is acceptable and still considered fair. 
If the network manager wishes that all \iaps~should try to capture perfectly equal amounts of bottleneck bandwidth, then they must set $C_\tau=0$.

The term underlined with double lines enables the DRL model to continuously improve the value $o^i$ of the objective by comparing it with the weighted average $\overline{o^i}$.
The formulation of this double underlined term ensures that the current performance surpasses or temporally maintains past performance.
In this term, $\lambda$, is a buffer function to absorb fluctuations in the objective value to some extent because the fluctuations are indispensable in networked environments due to the non-static nature of the traffic.
This function is defined as follows:
\begin{equation*}
    \lambda\left\{x\right\} =\begin{cases} x-C_o & \text{if}\ x>C_o  \\
              x+C_o & \text{if}\ x<-C_o \\
              0      & \text{otherwise}
    \end{cases}
\end{equation*}
As long as the fluctuations in the objective value fall within the range of $(-C_o,C_o)$, the reward function discards them.
The two constants, $C_\tau$ and $C_o$, make our DRL model highly resilient to noises from a variety of sources in the network.

\subsection{Action Space}
\label{subsec:ActionSpace}
For any given \iap, at any given time step $t$, the inference-model that the CC-client has for that \iap~outputs a value of \texttt{cwnd} for use at time step $t+1$.
Another approach of determining the value of \texttt{cwnd} for the next time step would be to pre-define a step-size $\Delta$, use the model as a two class classification model, and add/subtract $\Delta$ from the current \texttt{cwnd} value based on the output of the model.
We went with the first approach because it makes the model more adaptive: the model is able to rapidly make large changes to \texttt{cwnd} or gradually make small changes to \texttt{cwnd} as needed in response to drastic or slow changes, respectively, in network conditions.

\subsection{Neural-Network Architecture}
\label{subsec:NetworkArchitecture}
The neural-network architecture used in our \approachName~framework is shown in Fig. \ref{networkArchitecture}.
This figure shows the architecture of both central-model as well as the sub-models.
As described in Sec. \ref{subsec:SystemState}, we use 12 attributes to quantify the system state of each \iap.
Before feeding these attributes to the neural-network shown in Fig. \ref{networkArchitecture}, the CC-server first reshapes them into a $12\times1$ vector.
\begin{figure}[b]
\vspace{-0.2in}
\centerline{
        \includegraphics[width=0.9\linewidth,keepaspectratio]{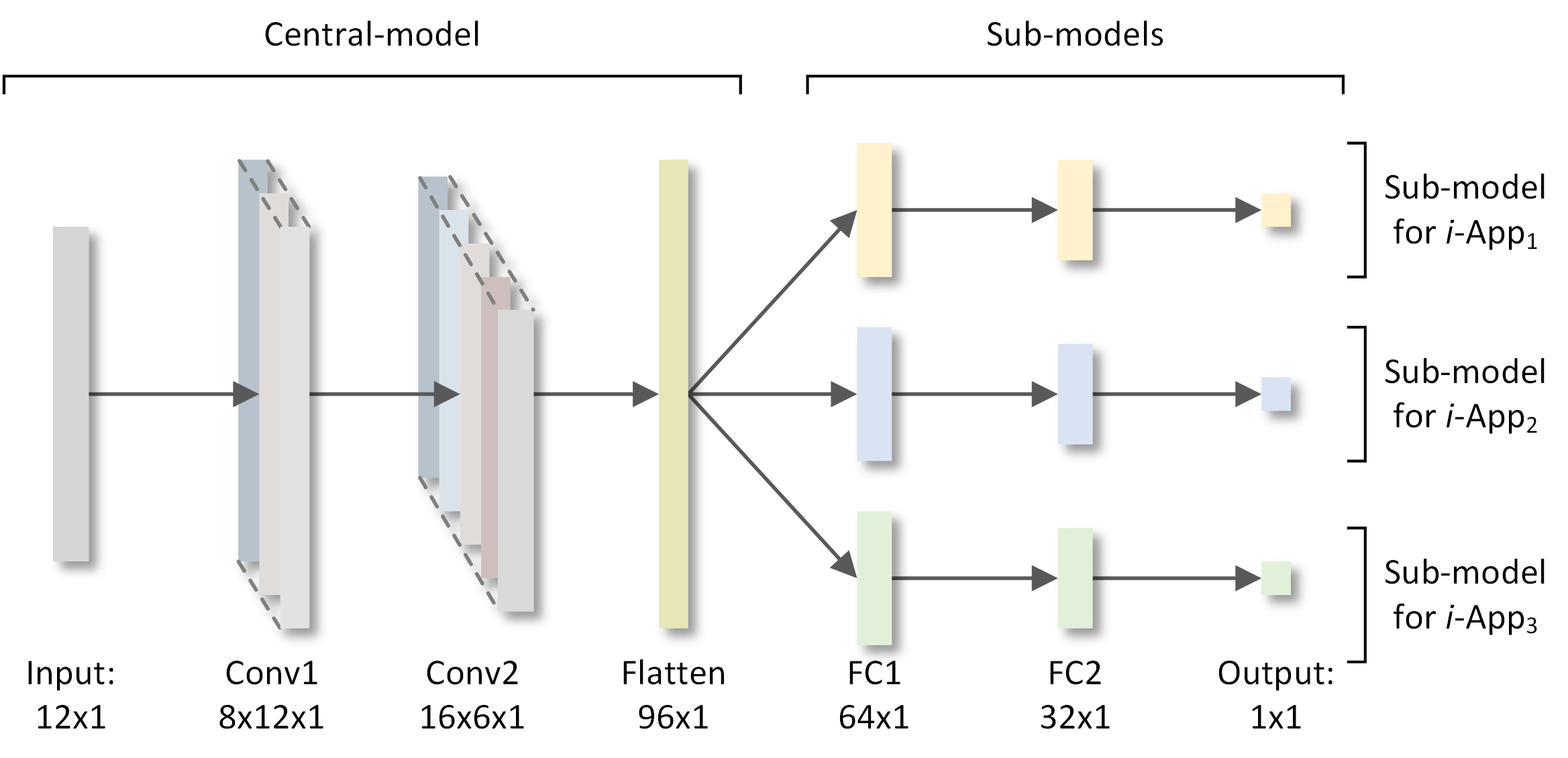}}
    \caption{Neural-network architecture}
    \label{networkArchitecture}
\end{figure}

The central-model acts as a feature extractor and its role is to derive a high-level representation of the system from raw system state.
This model encompasses 2 one-dimensional convolutional layers and 1 flattening operation.
In accordance with the \approachName~structure, the CC-server maintains a sub-model for each \iap.
Each sub-model comprises 2 dense hidden layers with shapes of 64x1 and 32x1, respectively.
We employ batch normalization, scaling the output of preceding layers to mitigate gradient vanishing, thereby significantly reducing training complexity of the neural-network.
The final output of this double-layered multi-layer perceptron is the value of \texttt{cwnd} that the CC-client uses in the next time step.

We used convolutional neural networks (CNN) instead of the fully connected networks in the main model as the feature extractor for two reasons.
First, CNNs excel in extracting spatial information \cite{kattenborn2021review,alzubaidi2021review}, which is crucial in our context to discern the inherent relationships between various attributes.
Second, the parameter count in CNNs is lower \cite{briot2018analysis,kim2017performance}, which makes them computationally feasible for use on CC-clients for inference.
%


\subsection{Fast Retraining Mechanism}
Recall that an \iap~has the liberty to change its optimization objective at any time.
Typically, this would necessitate a complete retraining of its sub-model.
We introduce fast retraining mechanism (FRM), aimed at substantially reducing the retraining latency.
FRM works by leveraging the observations that in a given non-wide area network, if an \iap~is running on one CC-client, it is quite likely that other instances of that \iap~are running on other, or even the same, CC-client(s).
Consequently, it is highly likely that some other instance(s) already uses the optimization objective that the current instance of the \iap~just switched to.
If it indeed is the case, the CC-server first identifies the corresponding existing sub-model for the \iap~instance with the same objective.
Next, it copies the 96$\times$64 weight matrix between Faltten and FC1 layers and the 64$\times$32 weight matrix between the FC1 and FC2 layers from the existing sub-model to the new sub-model.
By employing this strategy of initialization, \approachName~enables new sub-model to effectively leverage past knowledge, drastically reducing adaptation time compared to complete retraining of the sub-model from scratch.
Furthermore, we randomly initialize the 32$\times$1 weight matrix between the FC2 and the output layers of the new sub-model instead of copying it from an existing sub-model.
This enables the new sub-model to learn new information from the system state sent by its corresponding \iap~instead of just copying everything from another sub-model related to a different instance of that \iap.
In our current implementation, we do not rely on any form of pre-training on simulators.
Instead, the central-model accumulates knowledge over time and each new sub-model benefits from it through parameter sharing.
Bootstrapping from a pre-trained model may further enhance performance in early steps, and we leave that exploration to future work.

\subsection{Scalability}
An important property inherent in our design of \approachName~is scalability.
This scalability stems from the simplicity with which \approachName~can handle new as well as terminating TCP connections.
Whenever an \iap~establishes a new TCP connection, the CC-server simply starts a new sub-model for this \iap~and uses DRL to train over time.
This training is quite fast because the central-model already exists and incorporates the latest information about the current status of the network.
If the optimization objective of the new \iap~matches with that of another current \iap, the CC-server further employs FRM described above to speed up the training of the new sub-model for it to quickly start achieving that optimization objective.
Similarly, if an \iap~terminates its TCP connection, the CC-server simply discards its sub-model.

\section{Experiments}
In this section, we extensively evaluate \approachName~from five perspectives:
1) achieving the optimization objective,
%
2) impact and overhead of the updating interval,
%
3) robustness in dynamically changing environments, 
4) effectiveness of FRM in adapting to new objectives, and
5) fairness in utilizing network resources.
We conducted our evaluations on three networks with well-known topologies \cite{lebiednik2016survey}, namely dumbbell, leaf-spine, and fat-tree, shown in Fig. \ref{fig:topologies}.
The bandwidth of bottleneck links (thick black) in each topology was kept at 2Mbps.
All other links had the capacity of 10Mbps.
%
%
Each simulation experiment ran for 60,000 steps, each step spanning 50ms.
For every configuration, we conducted 10 independent runs and report aggregate statistics.
Application traffic followed periodic ON/OFF burst models.
Unless otherwise specified, the learning rate is set to \texttt{1e-4} and SAC is trained with a buffer size of \texttt{1e5}.

\begin{figure}[htbp]
    \centering
    \begin{subfigure}[b]{0.32\linewidth}
        \includegraphics[width=\linewidth]{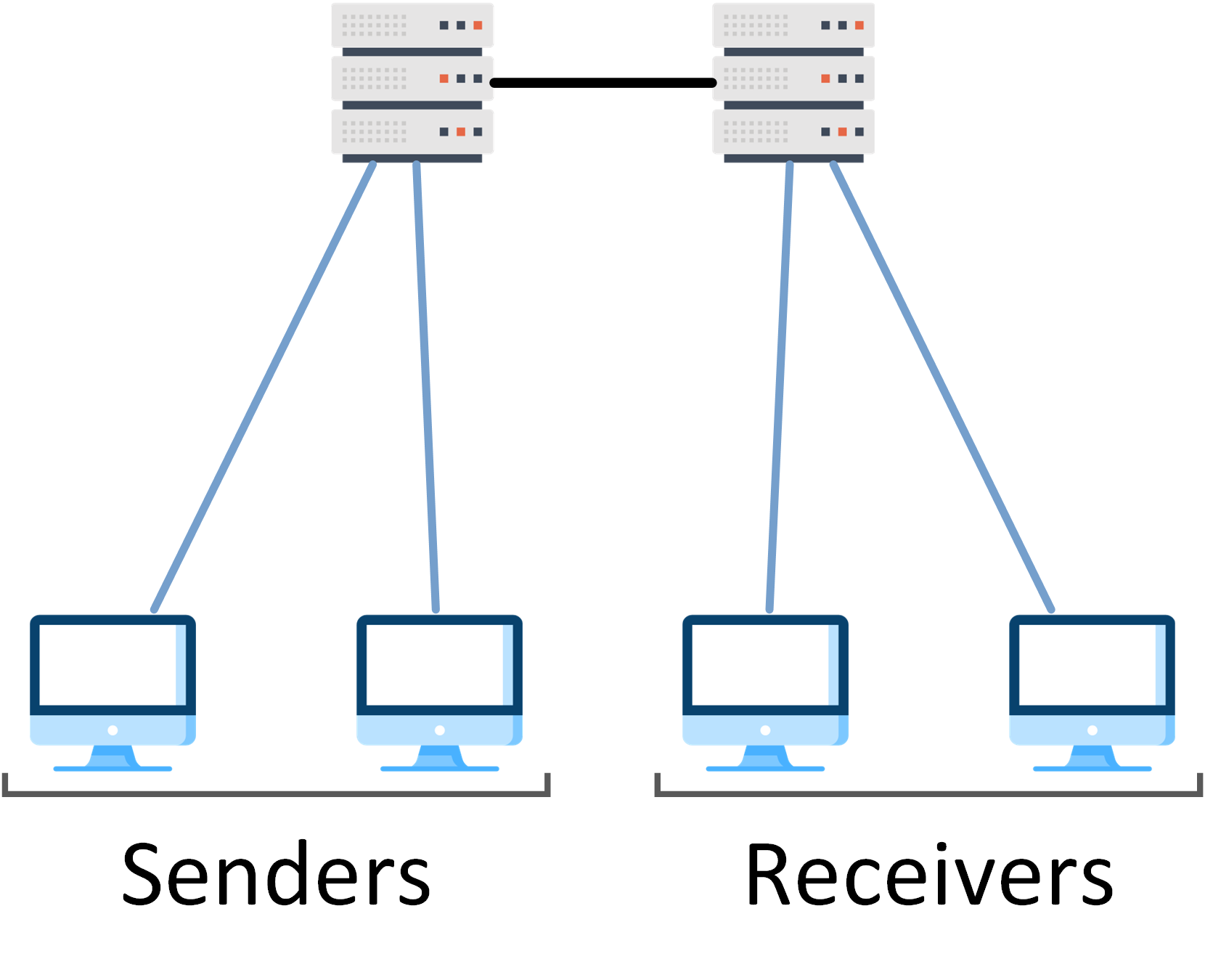}
        \caption{Dumbbell}
        \label{fig:dumbbelltopFig}
    \end{subfigure}
    \hfill
    \begin{subfigure}[b]{0.32\linewidth}
        \includegraphics[width=\linewidth]{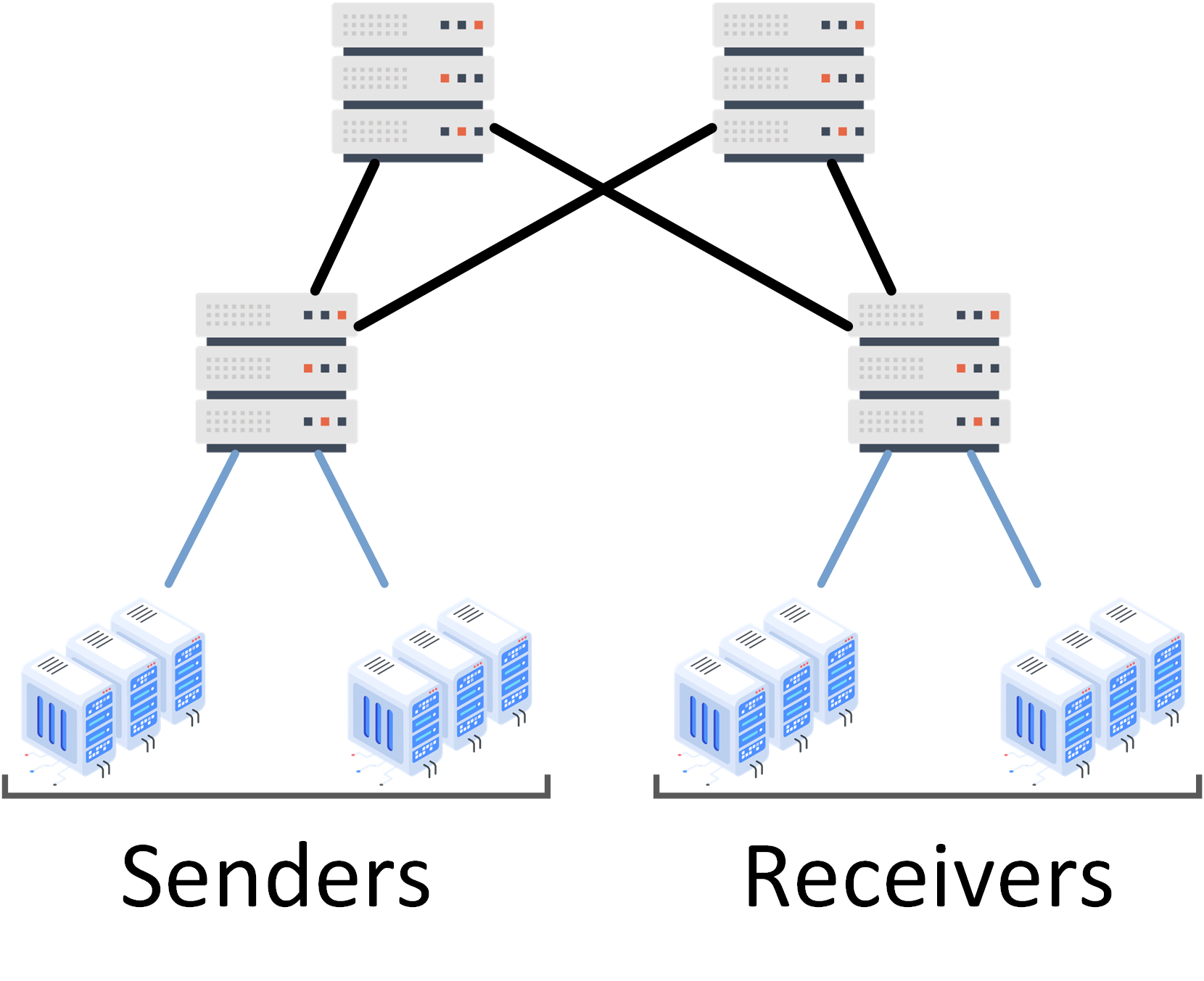}
        \caption{Leaf-Spine}
        \label{fig:LStopFig}
    \end{subfigure}
    \hfill
    \begin{subfigure}[b]{0.32\linewidth}
        \includegraphics[width=\linewidth]{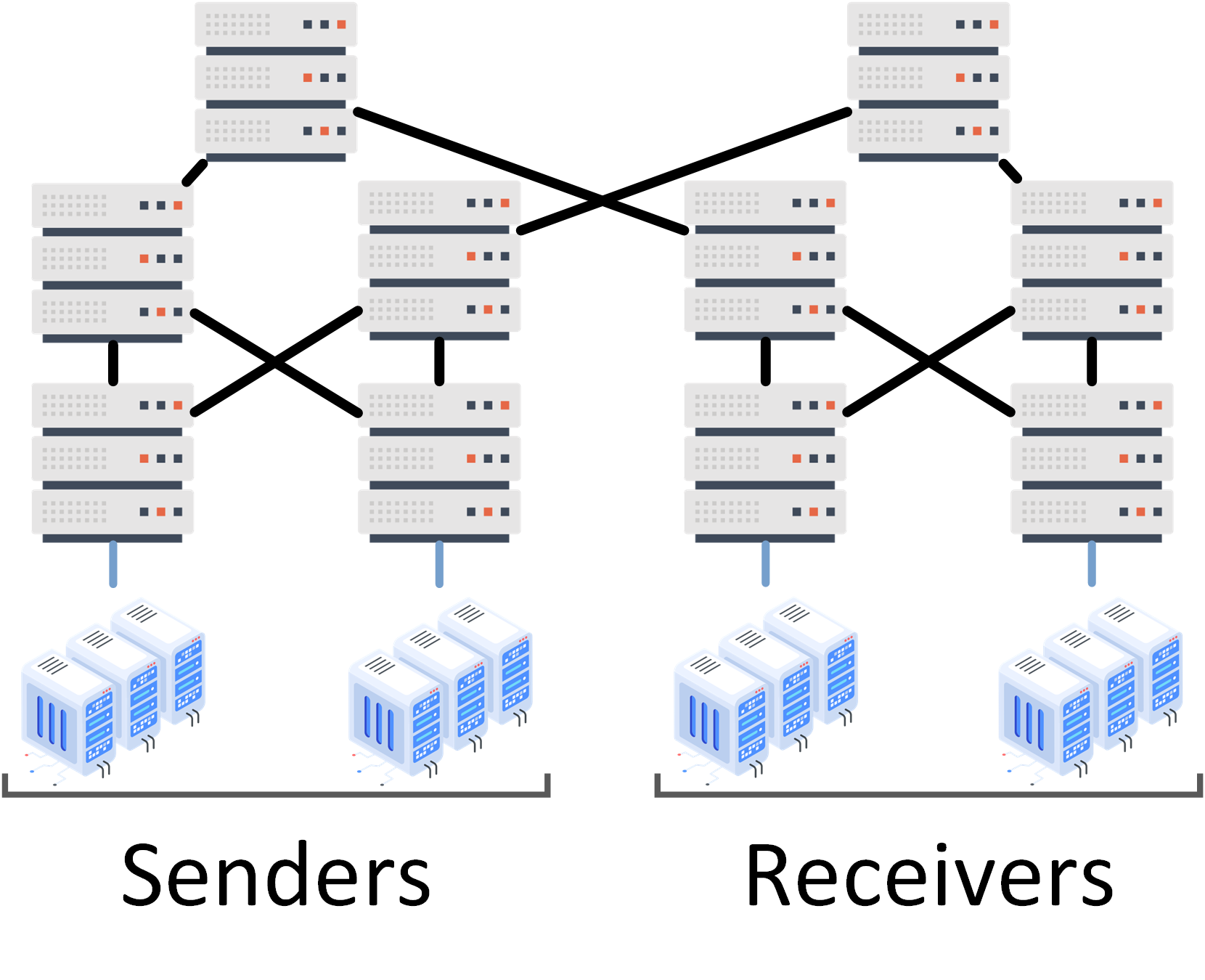}
        \caption{Fat-Tree}
        \label{fig:FTtopFig}
    \end{subfigure}
    \caption{Three topologies used in evaluations.}
        \label{fig:topologies}
\end{figure}

In addition to measuring the performance of \approachName, we further compared it with prior approaches.
For this, we selected two recent DRL-based algorithms, TCPAI \cite{yin2020ns3} and PCCRL \cite{Tessler2022} as well as two heuristic-based algorithms, NewReno \cite{Abdeljaouad2010} and Cubic \cite{Ha2008}.
%
%
We emphasize here that while we compare \approachName~with TCPAI and PCCRL, both TCPAI and PCCRL suffer from all the limitations mentioned in Sec. \ref{subsec:LimitationsofPriorArt}.
In particular, note that they can only achieve a single predefined optimization goal, \ie, \iaps~ cannot set their own customized optimization objectives.
Next, we present observations from our experiments conducted in \texttt{ns}-3 \cite{henderson2008network} on the five perspectives listed above.

%

\subsection{Achieving the Objective}
\label{subsec:AchievingtheObjective}
%
%
While an \iap~ can specify the optimization goal in \approachName~using any network metric, for the purpose of these experiments, we worked with three metrics: throughput, latency, and jitter.
Fig. \ref{fig:throughputObjective} shows the throughput achieved by the five approaches in each of our three topologies when the objective set for \approachName~was to maximize throughput.
Each throughput is shown as box-plot to highlight the variation in throughput exhibited by different approaches.
To make the box-plot of throughput for any given approach in a given topology, we calculate the average throughput achieved by each sender node in the topology during each updating interval, and then take all these values from all senders and make a box-plot from them. 
%
%
As existing approaches do not have a notion of an update interval, for any existing approach, we use the value of the update interval used in \approachName~as the duration over which we calculate the average values of the throughput of the existing approach as well.
We observe from Fig. \ref{fig:throughputObjective} that \approachName~achieves the highest throughput in all topologies.
We further observe that \approachName~also exhibits the lowest variance in the throughput, which shows that \approachName~not only achieves the highest but also the most stable throughput among all approaches. 
%


\begin{figure*}[htbp]
    \centering
    \begin{subfigure}[b]{0.31\linewidth}
        \includegraphics[width=\linewidth]{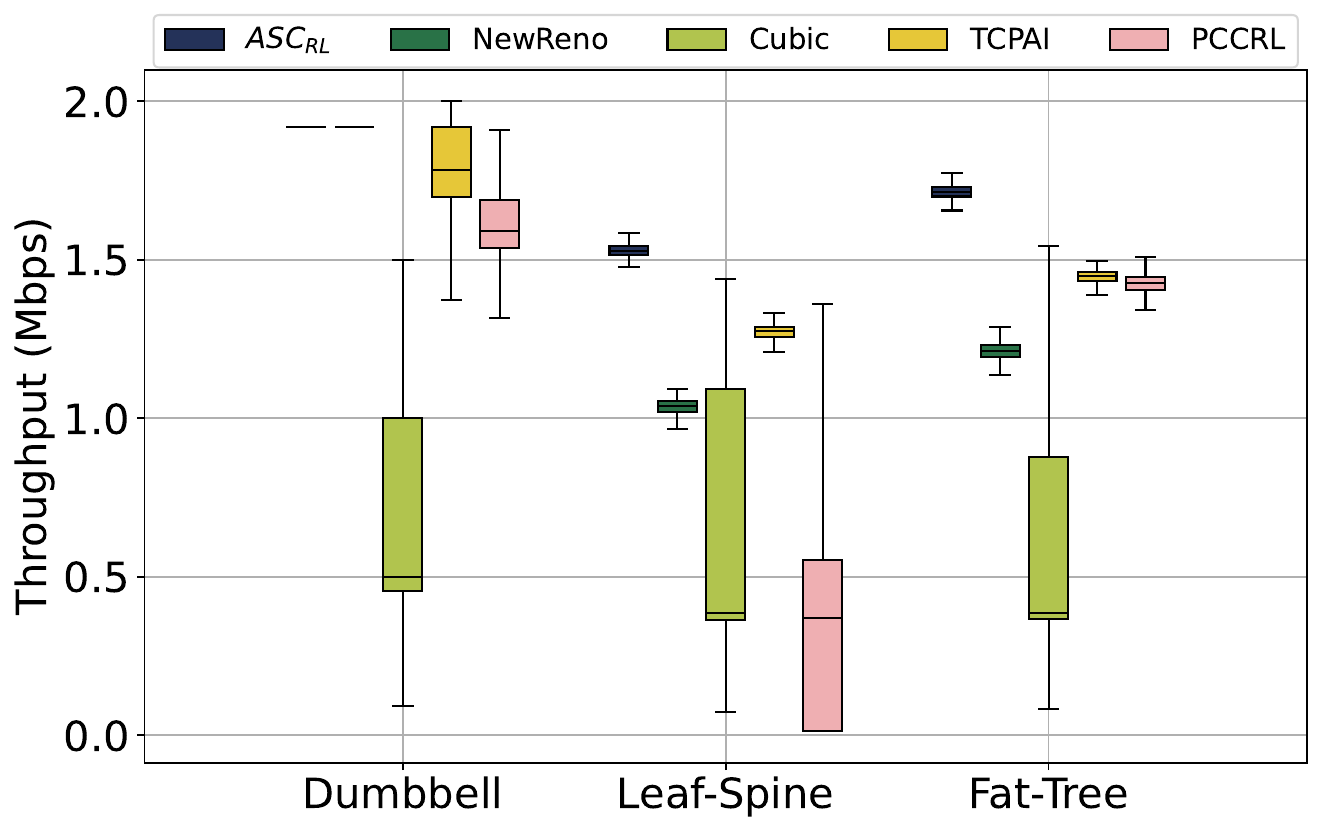}
        \caption{Maximize Throughput}
        \label{fig:throughputObjective}
    \end{subfigure}
    \hfill
    \begin{subfigure}[b]{0.317\linewidth}
        \includegraphics[width=\linewidth]{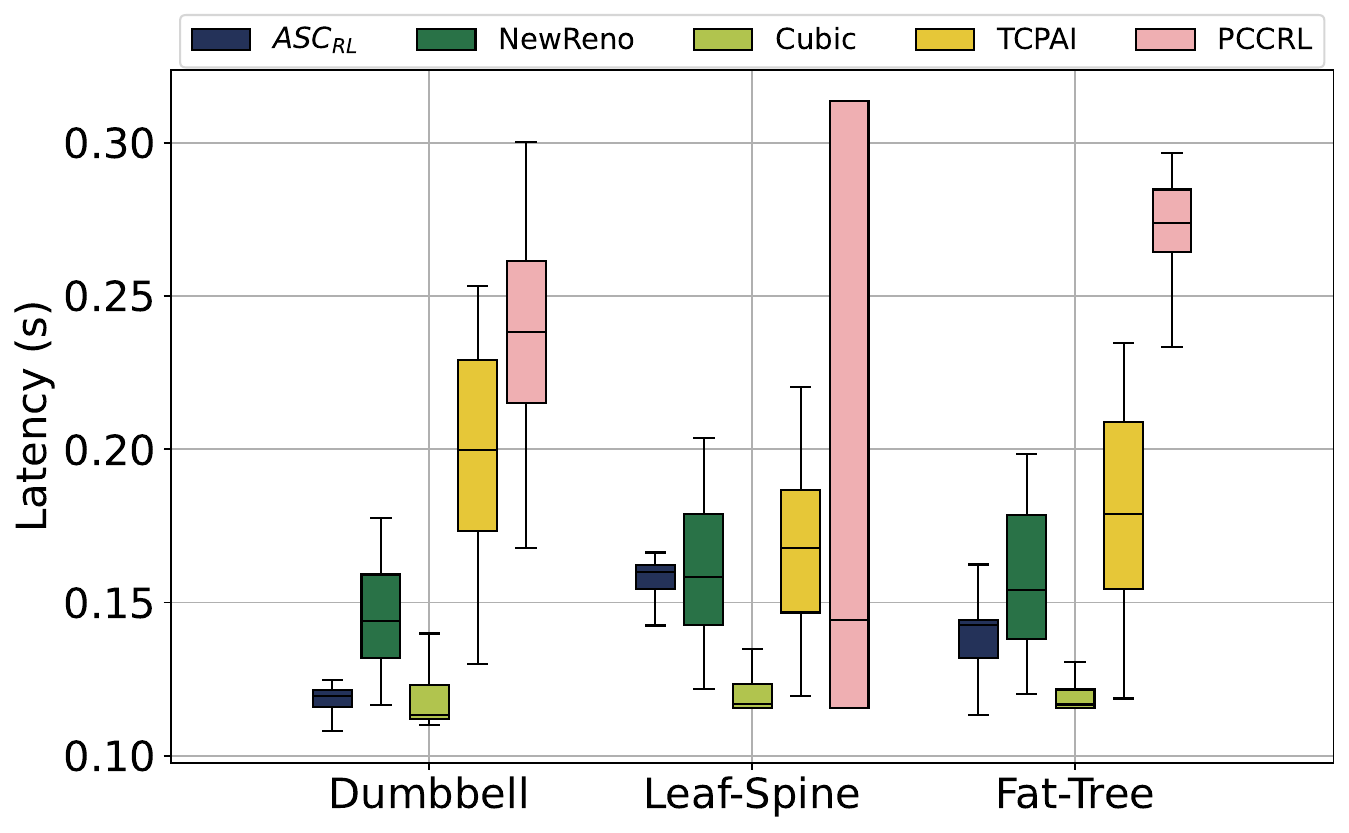}
        \caption{Minimize Latency}
        \label{fig:latencyObjective}
    \end{subfigure}
    \hfill
    \begin{subfigure}[b]{0.345\linewidth}
        \includegraphics[width=\linewidth]{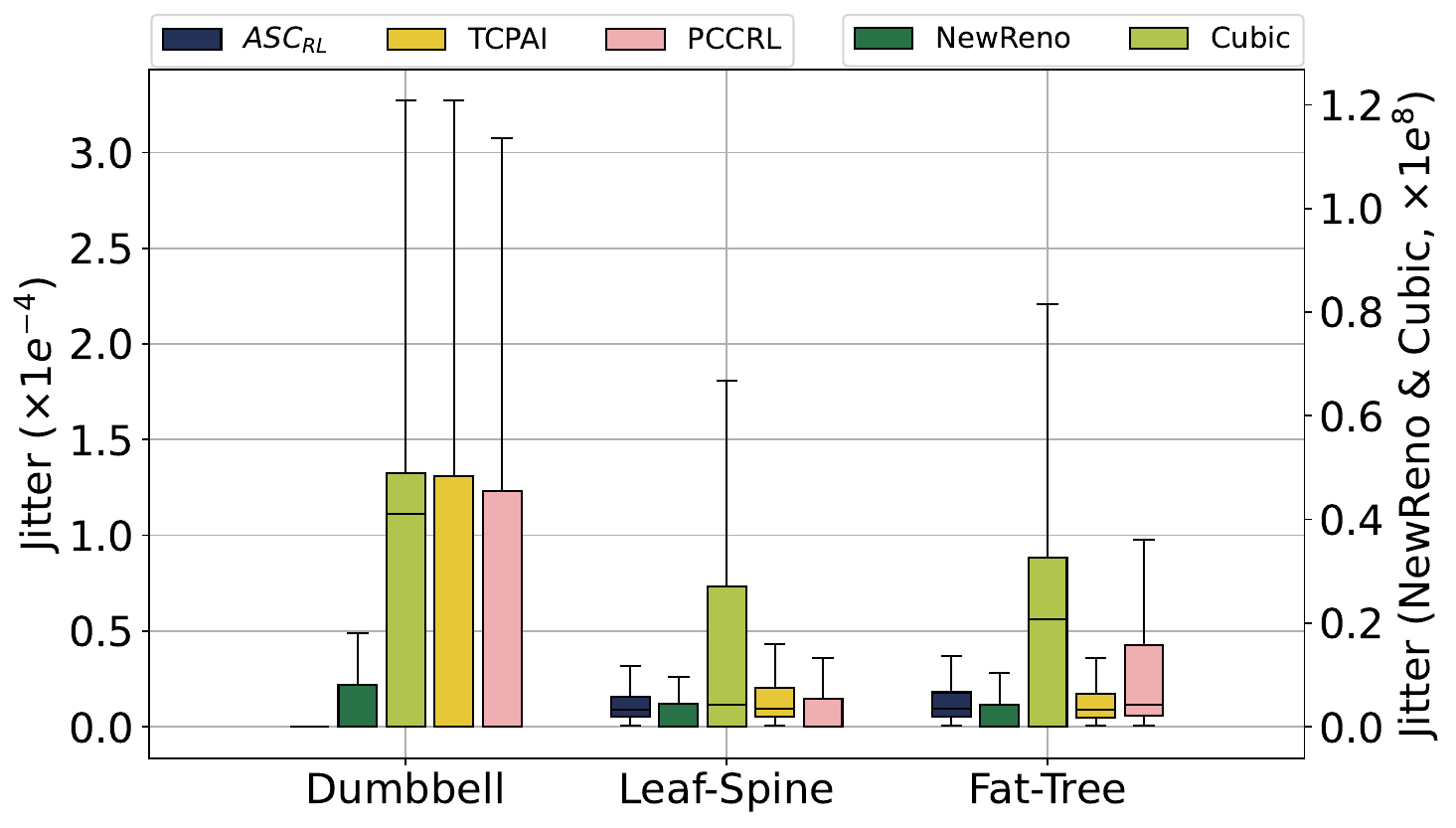}
        \caption{Minimize Jitter}
        \label{fig:jitterObjective}
    \end{subfigure}
    \caption{Performance of \approachName~and prior approaches when the objective was set as listed under each figure.}
    \label{convergence of three goals under dumbbell topology}
\end{figure*}

Figs. \ref{fig:latencyObjective} and \ref{fig:jitterObjective} show latency and jitter, respectively, experienced by the five approaches in each of our three topologies when the objective we set for \approachName~was to minimize latency and to minimize jitter, respectively.
Again, \approachName~ outperforms all other approaches, except CUBIC when optimizing latency under Leaf-Spine and Fat-Tree topologies.
However, note from Fig. \ref{fig:throughputObjective} that CUBIC achieved the lowest throughput, \ie, it keeps such little data in-flight that its packets experience hardly any congestion, and thus see the lowest latency.
While we saw in Fig. \ref{fig:throughputObjective} that the throughput's achieved by some of the other approaches were close to \approachName's, we see a significantly better performance achieved by \approachName~ in latency and jitter compared to the other approaches. 
%
%
In fact, for New Reno and Cubic, we had to employ a secondary y-axis due to their very high jitter compared to \approachName.
%
%



\subsection{Impact and Overhead of Updating Interval}
%
The duration between sending the successive system states to the CC-server, \ie, the updating interval, can significantly influence the performance of \approachName.
If the updating interval is too small, it would lead to increased computing and network overhead as well as partial-information and biased data.
This results in less accurate sub-models, which in turn hinder in achieving the desired objective.
If the updating interval is too large, it would prolong the convergence time, again resulting in sub-optimal performance due to the lack of prompt environmental awareness.

Fig. \ref{fig:dumbellthpt} plots the cumulative distribution function (CDF) of throughput across different experiments in the dumbbell topology when the optimization goal was to maximize throughput, where in each experiment we used a different updating interval.
Fig. \ref{fig:dumbelllat} plots the same for latency.
We have not included the figure for jitter due to space contraint and because the observations were largely similar.
We observe from Fig. \ref{fig:dumbellthpt} that \approachName~ achieves the best throughput performance when each CC-client sends information about the system-state as soon as each of the \iap~ running has sent 100 packets (assuming each \iap~ always has data to send).
Recall from Sec. \ref{subsec:SystemState} that system-state is comprised of 12 values, each of 4 bytes.
Assuming a 2 byte \iap~ identifier (which could simply be the TCP port number), a typical 1500-byte Ethernet/WiFi packet can carry system state of over 25 \iaps~.
For a typical CC-client that has 10 \iaps~ with active TCP connections, the network overhead introduced by the use of \approachName~ comes out to be just 0.1\%.
%
\begin{figure}[htbp]
    \centering
    \begin{subfigure}[b]{0.49\linewidth}
        \includegraphics[width=\linewidth]{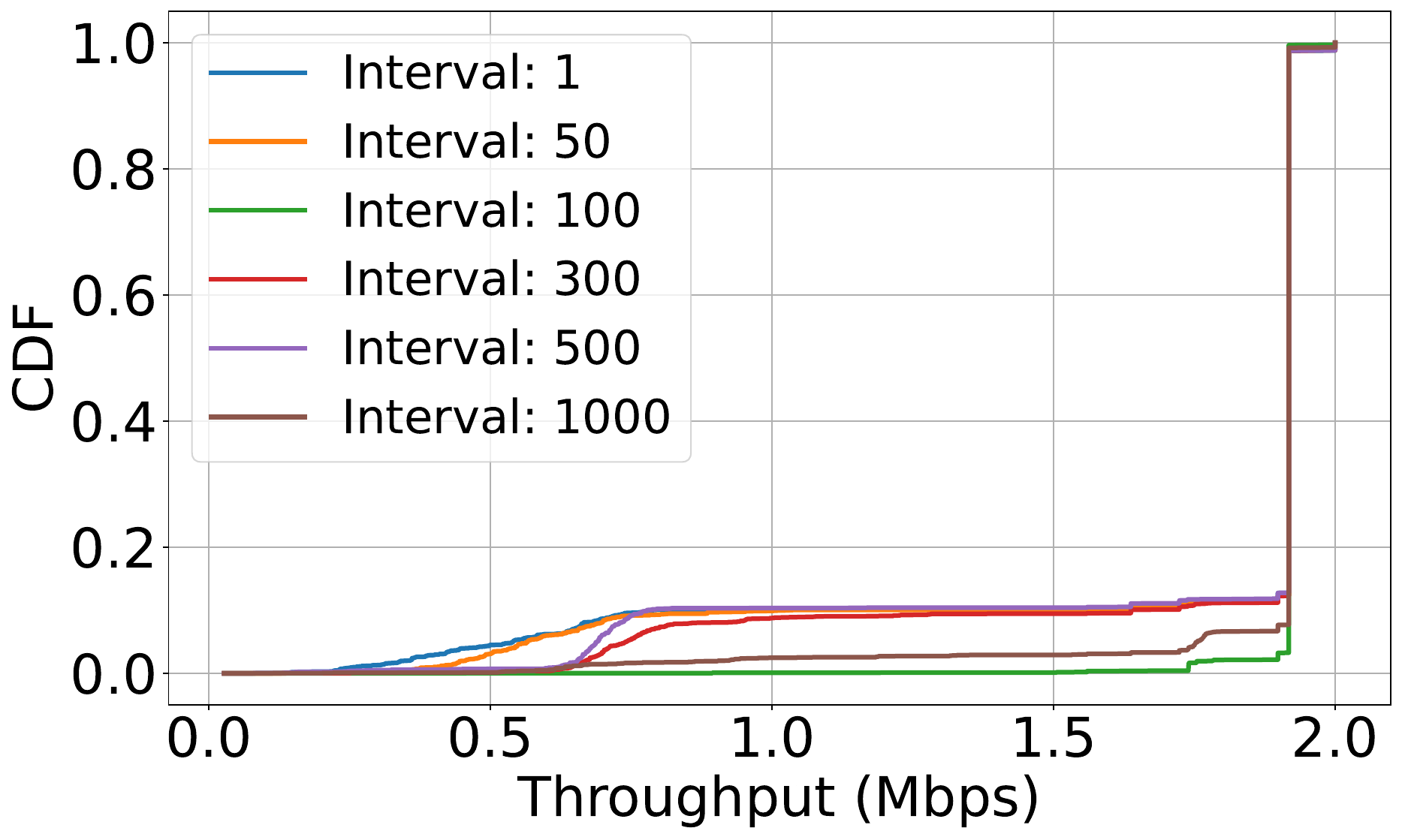}
        \caption{Throughput}
        \label{fig:dumbellthpt}
    \end{subfigure}
    \hfill
    \begin{subfigure}[b]{0.49\linewidth}
        \includegraphics[width=\linewidth]{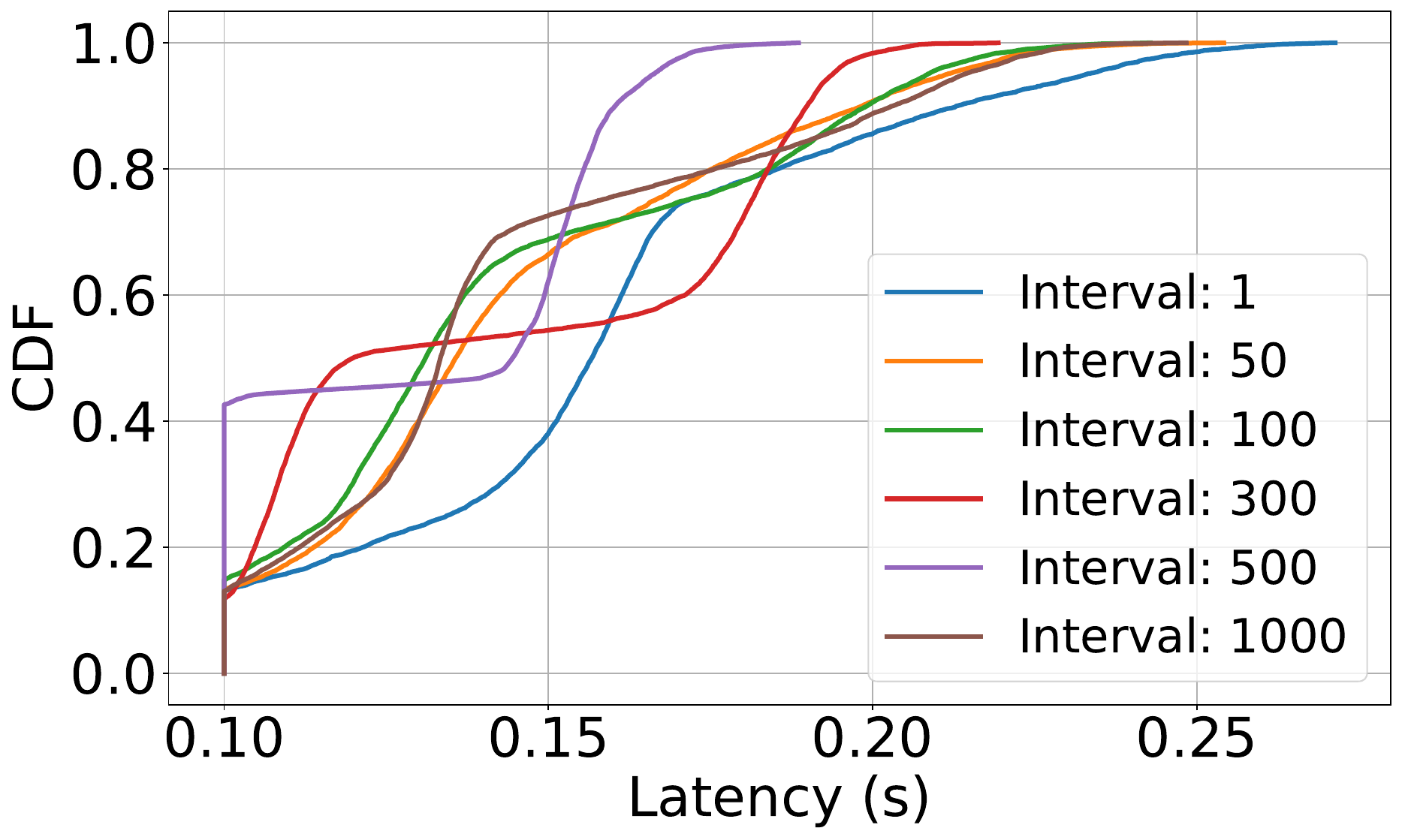}
        \caption{Latency}
        \label{fig:dumbelllat}
    \end{subfigure}
   \caption{CDFs of throughput and latency for different updating intervals in dumbbell topology}
    \label{fig:CDFDumbbell}
\end{figure}

We further observe from Fig. \ref{fig:dumbellthpt} that smaller updating intervals, such as 1, as well as larger updating intervals, such as 1000, perform worse compared to the updating interval of 100.
This is because smaller updating intervals provide instantaneous view of the network, which may not be representative of the overall current performance of the network.
Similarly, while the larger updating intervals mitigate biases introduced by the instantaneous view, they delay environmental awareness, resulting in poorer performance due to longer convergence times.
In the dumbbell topology, the interval of 100 strikes a balance between timely information acquisition and contained data bias.

We make similar observations from Fig. \ref{fig:dumbelllat}, where the interval of 500 packets performed best, and the latency gradually worsened (increased) as the updating interval decreased from 500 to 1 and also as it increased from 500 to 1000.
This again underscores the importance of a moderate update interval in reducing bias and maintaining prompt environmental awareness.
We made very similar observations from fat-tree as well as leaf-spine topologies, except that the value of the optimal update intervals turned out to be larger in these two topologies.
This happened because the complexity of these topologies demand more data to effectively eliminate any biases introduced by short-term observations.

All these observations collectively highlight that there is no single value of updating interval that is optimal for all settings.
The pivotal aspect lies in the versatility afforded by our proposed framework, offering users the flexibility to select appropriate updating intervals based on client resources and specific optimization goals.
Another way of setting the updating interval would be where the CC-server continuously monitors the performance achieved by various \iaps~ on all the CC-clients and correlates that performance with their corresponding update intervals.
Over time, the CC-server can learn from this information and then periodically start estimating the optimal update interval for each \iap~ and informing the corresponding CC-clients about it.
%
We emphasize however, that this automatic estimation of the optimal update interval is beyond the scope of this paper: it is a complete project in itself, and will be the topic of our future work.

\begin{figure*}[htbp]
    \centering
    \begin{subfigure}[b]{0.32\linewidth}
        \includegraphics[width=\linewidth]{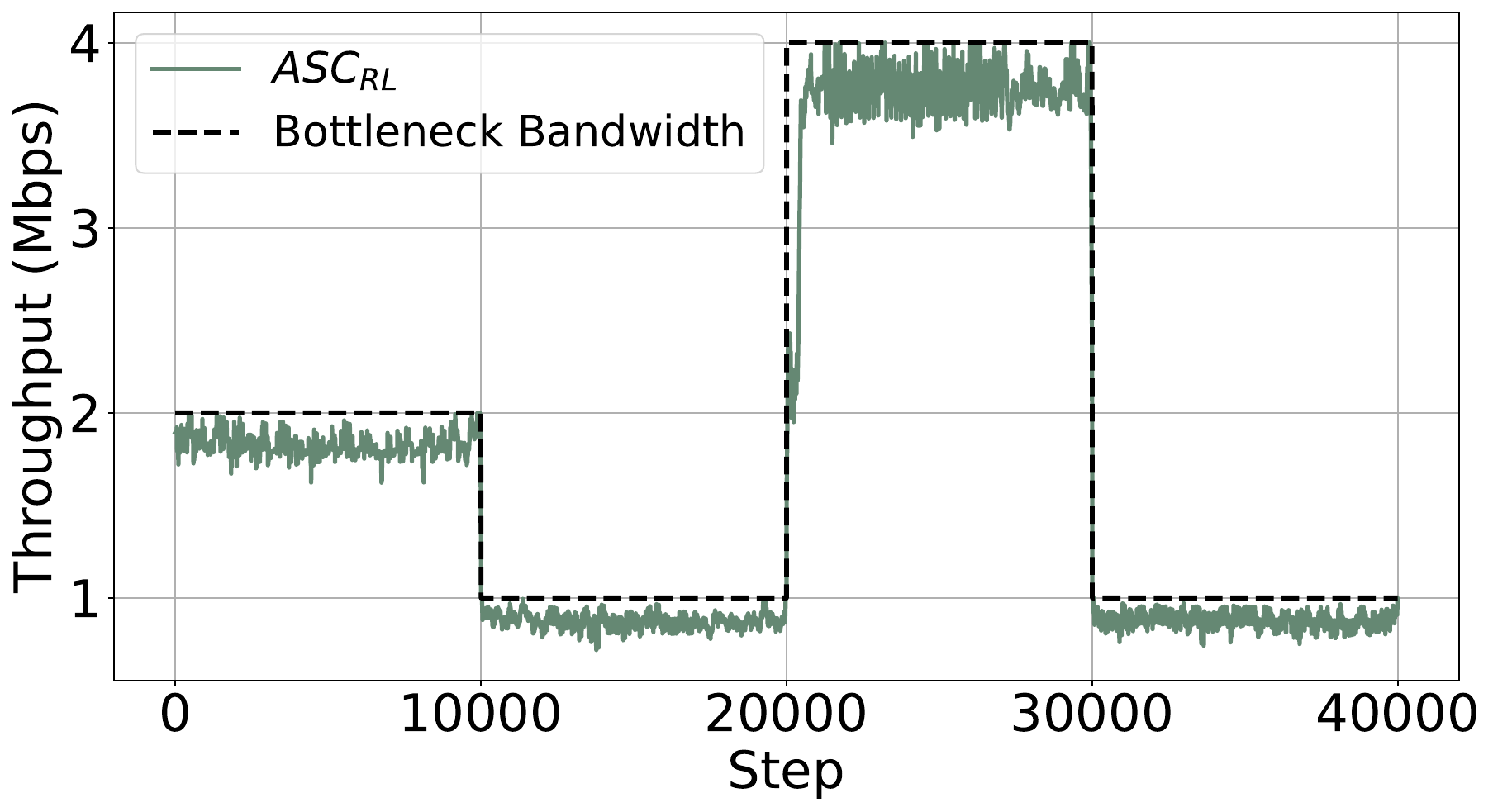}
        \caption{Dumbbell Topology}
        \label{fig:robustnessDumbbell}
    \end{subfigure}
    \hfill
    \begin{subfigure}[b]{0.32\linewidth}
        \includegraphics[width=\linewidth]{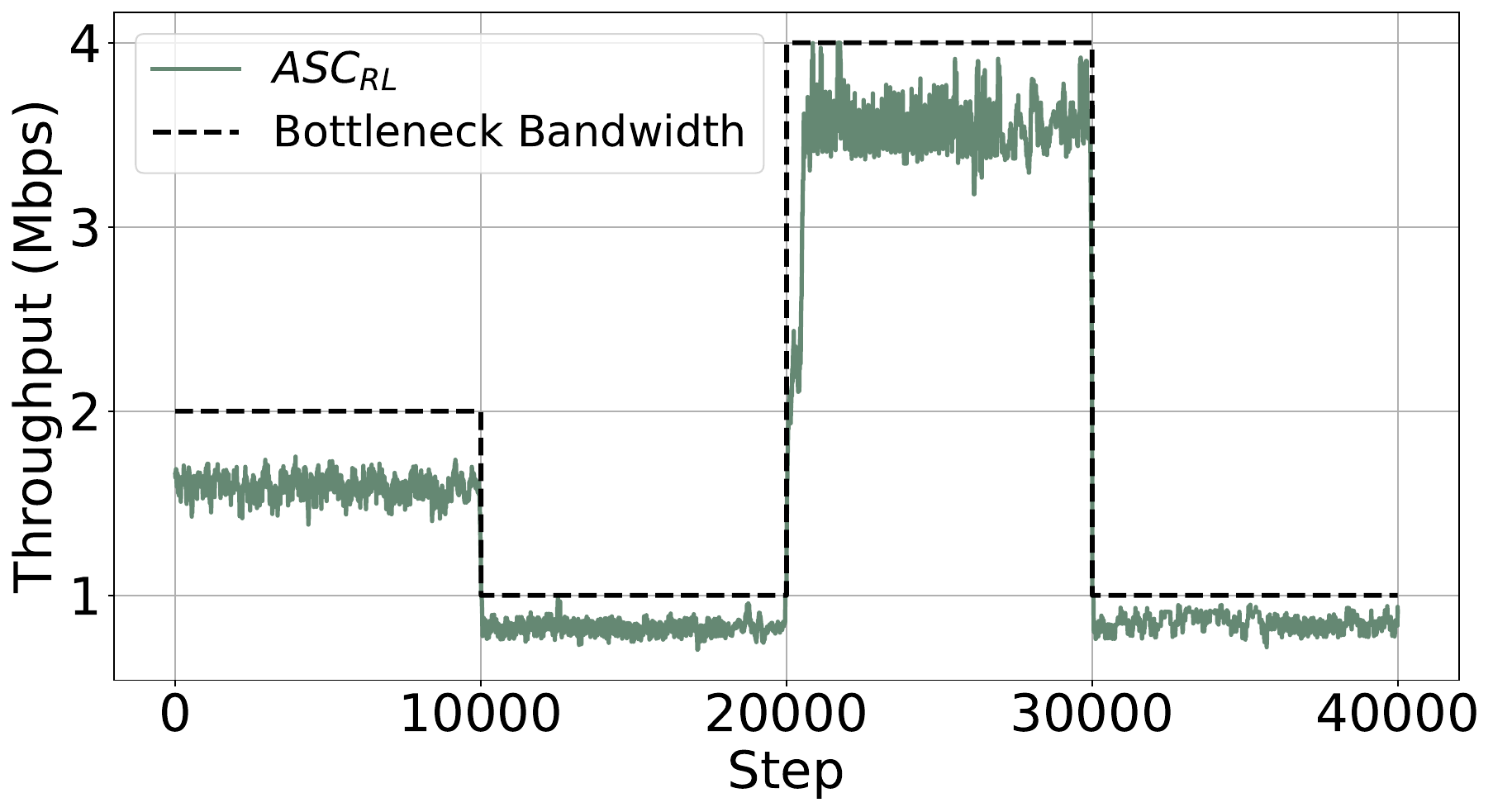}
        \caption{Leaf-Spine Topology}
        \label{fig:robustnessSpineLeaf}
    \end{subfigure}
    \hfill
    \begin{subfigure}[b]{0.32\linewidth}
        \includegraphics[width=\linewidth]{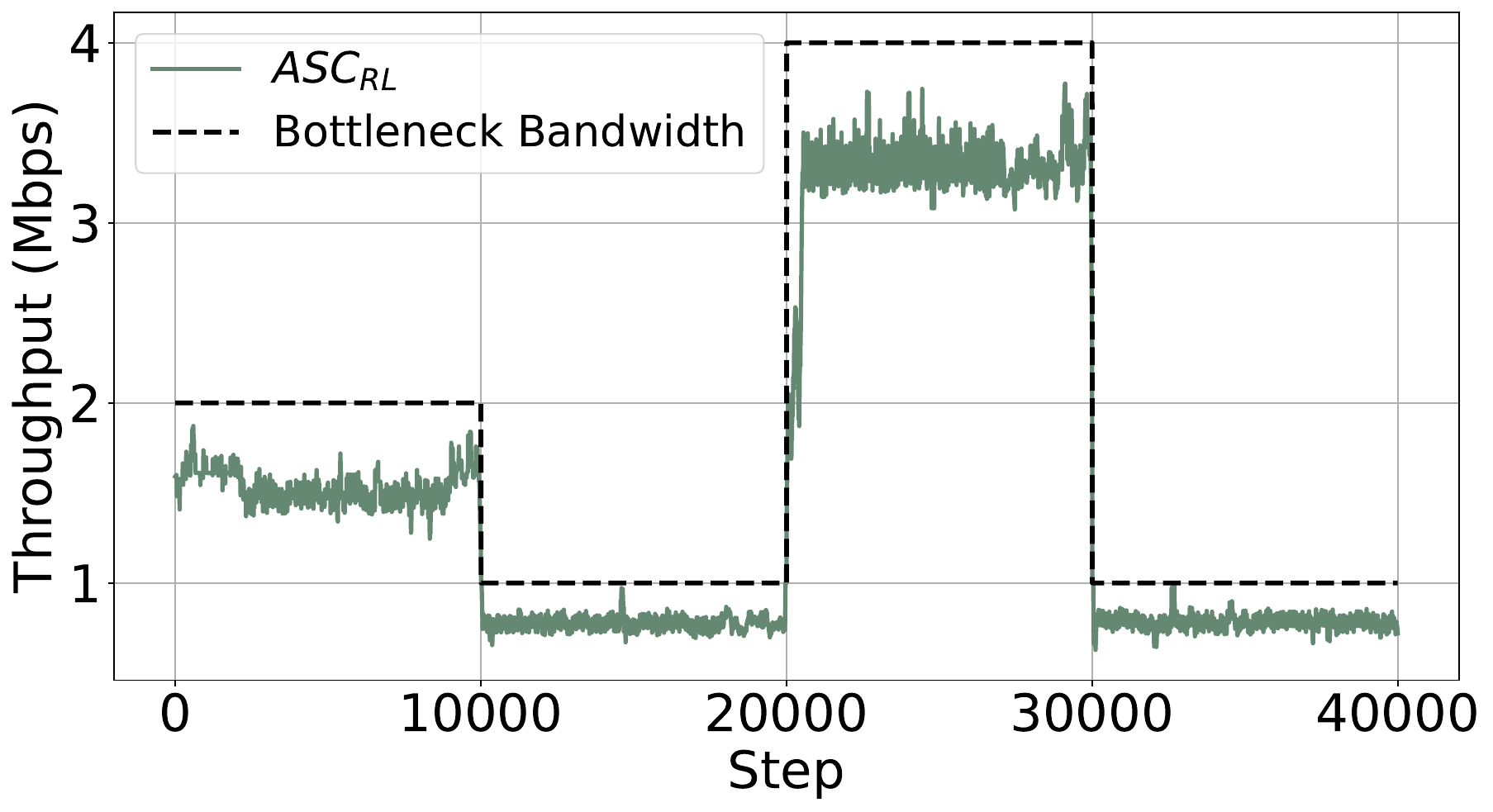}
        \caption{Fat-Tree Topology}
        \label{fig:robustnessFatTree}
    \end{subfigure}
    \caption{Robustness of \approachName~ in changing network conditions}
    \label{fig:robustness}
\end{figure*}

\subsection{Robustness in Dynamic Environments}
To study how quickly \approachName~ can learn about and adapt to changes in network conditions, we conducted experiments where we changed the bandwidth of bottleneck link multiple times, transitioning from 2Mbps to 1Mbps, then to 4Mbps, and finally back to 1Mbps, each bandwidth lasting for 10,000 DRL steps.
We set the optimization objective to maximize throughput.
Fig. \ref{fig:robustness} shows the total throughput achieved by the sender nodes in the three topologies.
We can see from this figure that \approachName~ framework swiftly adjusts to the new environment within just a few steps.
When the bottleneck bandwidth increases or reduces, the framework promptly increases or reduces, respectively, the congestion window size while aiming to achieve the maximum possible throughput.
Notice that the total throughput achieved in the leaf-spine and fat-tree typologies are slightly lower compared to the available bottleneck bandwidth.
This happens due to more potential packet loss because of multiple congestion points in the network.
%
Recall from Sec. \ref{subsec:AchievingtheObjective} that \approachName~ still outperforms all prior approaches, and at the same time, enables \iaps~ to set their own unique optimization objectives, a capability that existing approaches do not provide.
Nonetheless, the observations from Fig. \ref{fig:robustness} showcase \approachName's rapid environmental awareness and swift adaptation to dynamic changes, which makes it well-suited for deployment in dynamic and diverse network environments.
We made similar observations for several other optimization goals.

\subsection{Effectiveness of Fast Retraining Method}
Traditional DRL-based methods struggle with adapting to new optimization goals due to their specialized loss functions and rigid structures.
\approachName~ solves this challenge through its central-model+sub-model structure design along with FRM.
To validate its efficacy, we compare our approach to the conventional method of complete retraining from scratch, which prior approaches would have to do if the optimization goal of an \iap~ changes.
We examined three different transitions of optimization goals: from latency to throughput, from jitter to latency, and from throughput to jitter.
Fig. \ref{fig:goalTransLF} shows the results on fat-tree topology.
The observations on the other two topologies were very similar, and thus, we have not included those figures.

\begin{figure}[htbp]
    \centering
    \begin{subfigure}[b]{0.49\linewidth}
        \includegraphics[width=\linewidth]{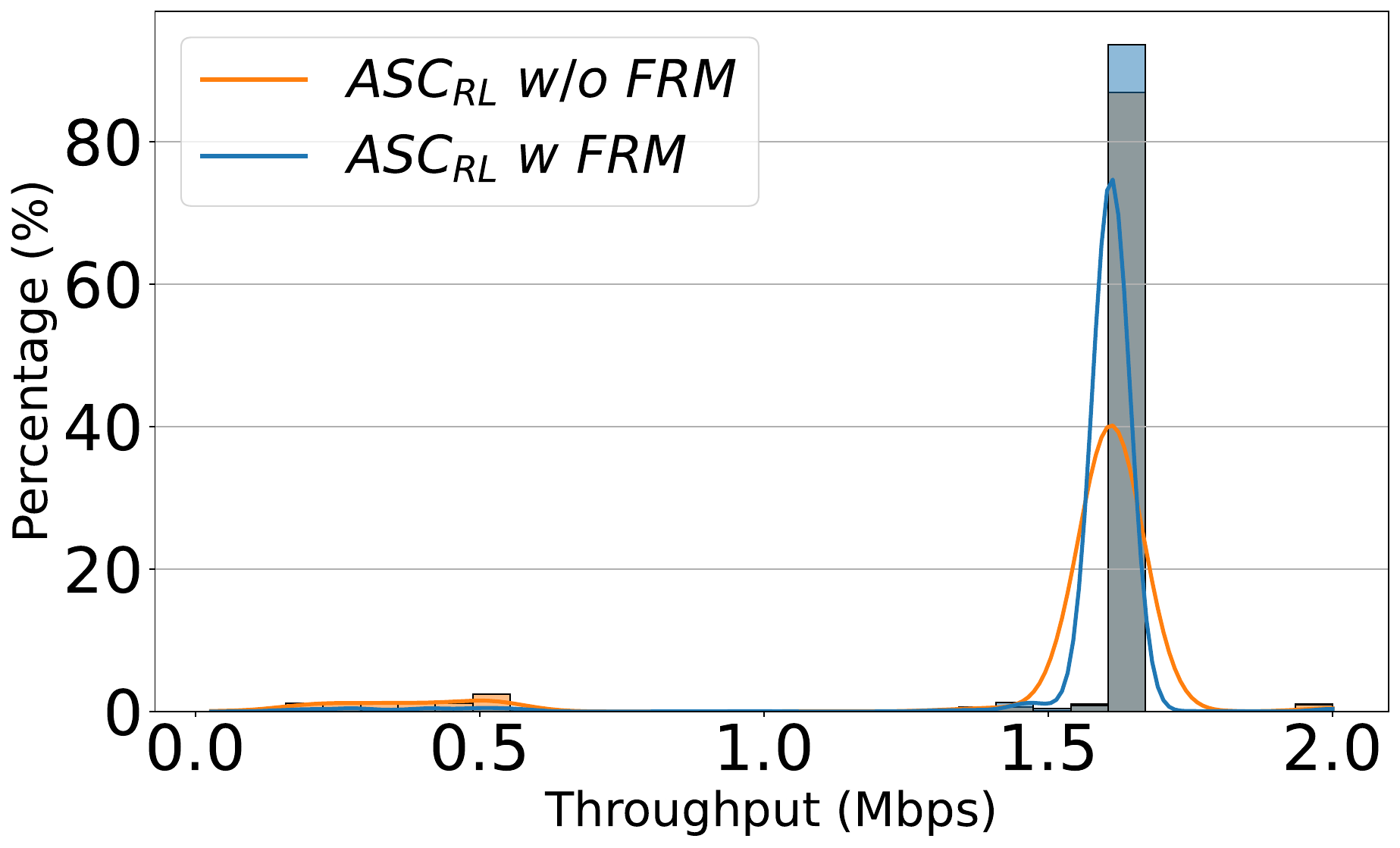}
        \caption{Latency to Throughput}
        \label{fig:goalL2TLF}
    \end{subfigure}
    \hfill
    \begin{subfigure}[b]{0.49\linewidth}
        \includegraphics[width=\linewidth]{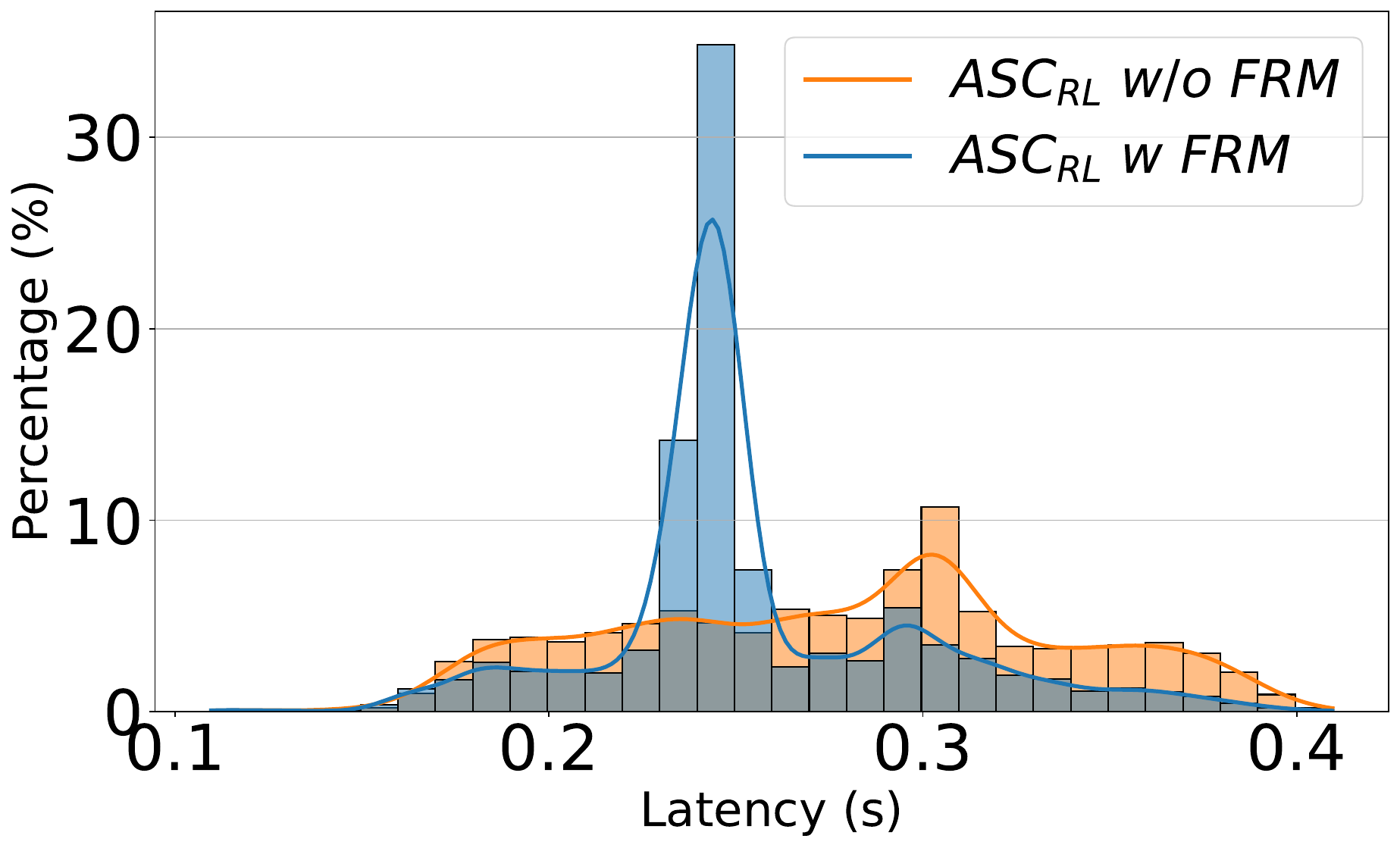}
        \caption{Jitter to Latency}
        \label{fig:goalJ2LLF}
    \end{subfigure}
    \caption{Performance when transitioning the optimization goal on Fat-Tree topology.}
    \label{fig:goalTransLF}
\end{figure}

Fig. \ref{fig:goalL2TLF} shows the distribution of throughput using histograms with kernel density estimations when the goal transitioned from latency to throughput.
Fig. \ref{fig:goalJ2LLF} similarly shows the distribution of latency when the goal transitioned from jitter to latency.
We omitted the figure for transition from throughput to jitter because both methods maintained jitter near 0.
We observe from Figs. \ref{fig:goalL2TLF} and \ref{fig:goalJ2LLF} that the use of FRM consistently outperforms the approach of retraining from scratch, surpassing it by 5\% in the latency-to-throughput scenario and by over 50\% in the jitter-to-latency scenario.
We continued the experiment for 10,000 DRL steps after the transition of goal.
Had we allowed it to continue for a longer duration, the performance of retraining from scratch would eventually approach that of FRM.
This shows that FRM is highly effectively in utilizing stored knowledge to swiflty achieve the new optimization objective.

\subsection{Fairness}
A fundamental aspect to take into consideration while designing a TCP congestion control algorithm is to ensure that multiple TCP connections using that algorithm must share the available resources equally.
For example, one TCP connection should not achieve significantly higher throughput compared to others by capturing a disproportionately large portion of the bottleneck bandwidth.
Therefore, in this section, we study that in settings where multiple \iaps~ intend to achieve the same objective, how fairly do the \iaps~ get access to resources when using the congestion control approach proposed in the \approachName~ framework.
%
To quantify the fairness, we calculated the Jain's fairness index across all \iaps~ running on all the CC-clients whose traffic was traversing a bottlneck link.
Let $n$ represent the number of such \iaps.
Jain's fairness index would fall between $1/n$ (worst case, where a single \iap~ captures all bandwidth and the remaining $n-1$ \iaps~ get no bandwidth) and 1 (best case, where each \iap~ captures the same amount of bandwidth).
Table \ref{tab:fairness} lists the values of Jain's fairness index for the three topologies and the three optimization objectives.
We can see in this table that the Jain's fairness index is close to 1 for all three topologies and all three objectives.
This shows that \approachName~ is able to equitably share available resources among all \iaps~.
This is enabled by the reward function we have used in \approachName, which penalizes models with performance differences exceeding the threshold $C_\tau$ across \iaps~.
\vspace{-0.05in}
\begin{table}[htbp]
\small
    \centering
    \caption{Jain's fairness index achieved by \approachName}
    \begin{tabular}{cccc}
            \hline
                        & Maximize      & Minimize      & Minimize\\
                        & Throughput    & Latency       & Jitter\\
            \hline
            \hline
            Dumbbell    & 0.8926            & 0.9984           & 0.9519  \\
            Leaf-Spine  & 0.8795            & 0.9982           & 0.9245  \\
            Fat-Tree    & 0.9486            & 0.9978           & 0.9279  \\
            \hline
        \end{tabular}
    \label{tab:fairness}
\end{table}

\vspace{-0.05in}
\section{Conclusion}
\vspace{-0.05in}
In this paper, we presented \approachName, a DRL based framework for TCP congestion control that separates the deployment of learning and inference modules, making it computationally very simple for use on commodity devices.
The architecture of \approachName~inherently makes it highly scalable.
They key technical novelty of \approachName~lies in its ability to handle arbitrary optimization objectives from \iaps~ as well as its ability to handle changes in objectives of \iaps~ at runtime.
The key technical depth of \approachName~lies in its design of the reward function that not only ensures that \iaps~ achieve their desired objectives but also ensures that the traffic of the \iaps~ traversing the network demonstrates fair use of the available resources.
%
We have demonstrated these aspects extensively through experiments.
Our results further showed that \approachName~can not only achieve arbitrary objectives specified by \iaps, it actually outperformed all prior approaches even in the specific objectives they were designed to achieve.
We envision that the framework introduced in this paper will serve as the foundational stepping stone for the research community to build further on it and harness the potential that the learning-based approaches hold for TCP congestion control.
While our current implementation is focused on TCP, the underlying mechanisms of \approachName~can be extended to support other transport protocols such as QUIC, particularly due to QUIC’s user-space stack flexibility. We plan to investigate this in our future work.
}

\bibliographystyle{unsrt}
\bibliography{UUDRL}

\end{document}